\documentclass[twocolumn,showpacs,preprintnumbers,amsmath,amssymb,superscriptaddress]{revtex4}

\usepackage{graphicx}
\usepackage{dcolumn}
\usepackage{bm}

\newcommand{\bx}{\text{box}}
\newcommand{\cl}{\text{cl}}
\newcommand{\crit}{\text{c}}
\newcommand{\hc}{}

\begin{document}

\preprint{APS/123-QED}

\title{Dilute Wet Granulates: Nonequilibrium Dynamics and Structure Formation}

\author{Stephan Ulrich}
\email{ulrich@theorie.physik.uni-goettingen.de}
\affiliation{Universit\"at G\"ottingen, Institute of Theoretical Physics, Germany}%

\author{Timo Aspelmeier}%
\affiliation{
Max-Planck-Institut f\"ur Dynamik und Selbstorganisation, Dept.~Dynamics of Complex Fluids, G\"ottingen, Germany
}%

\author{Annette Zippelius}%
\affiliation{Universit\"at G\"ottingen, Institute of Theoretical Physics, Germany}%
\affiliation{
Max-Planck-Institut f\"ur Dynamik und Selbstorganisation, Dept.~Dynamics of Complex Fluids, G\"ottingen, Germany
}%

\author{Klaus Roeller}
\affiliation{
Max-Planck-Institut f\"ur Dynamik und Selbstorganisation, Dept.~Dynamics of Complex Fluids, G\"ottingen, Germany
}%

\author{Axel Fingerle}
\affiliation{
Max-Planck-Institut f\"ur Dynamik und Selbstorganisation, Dept.~Dynamics of Complex Fluids, G\"ottingen, Germany
}%

\author{Stephan Herminghaus}
\affiliation{
Max-Planck-Institut f\"ur Dynamik und Selbstorganisation, Dept.~Dynamics of Complex Fluids, G\"ottingen, Germany
}%

\date{\today}

\begin{abstract}
  We investigate a gas of wet granular particles, covered by a thin
  liquid film. The dynamic evolution is governed by two-particle
  interactions, which are mainly due to interfacial forces in contrast
  to dry granular gases. When two wet grains collide, a capillary
  bridge is formed and stays intact up to a certain distance of
  withdrawal when the bridge ruptures, dissipating a fixed amount of
  energy. 
  A freely cooling system is shown to undergo a nonequillibrium dynamic
  phase transition from a state with mainly single particles and fast
  cooling to a state with growing aggregates, such that bridge rupture
  becomes a rare event and cooling is slow. In the early stage of
  cluster growth, aggregation is a self-similar process with a fractal
  dimension of the aggregates approximately equal to $D_\text{f}
  \approx 2$. At later times, a percolating cluster is observed which
  ultimately absorbs all the particles. The final cluster is compact
  on large length scales, but fractal with $D_\text{f} \approx 2$ on
  small length scales.
\end{abstract}

\maketitle

\section{Introduction\label{sec:Introduction}}

Granular materials are systems of macroscopic particles which interact
only when they are in mutual contact, and the interaction is
dissipative. In spite of this simple definition, collective phenomena
arising in such sytems are of utmost complexity, and have inspired
strongly increasing research activities in recent years.  The
particular interest in granular systems is mainly due to the fact that
their importance spans from technology and applied research to very
fundamental questions of interdisciplinary relevance. On the one hand,
storage and handling of bulk solids is among the most significant
tasks in industrial technology, and still poses a large number of
unsolved problems \cite{Jaeger96,deGennes1999,Duran00}. On the other
hand, granular systems provide a comparatively simple, experimentally
accessible model for physics far from equilibrium
\cite{PoeschelBuch,kudrolli2004,umbanhowar1996}. This is at the heart
of self-organization and pattern formation processes, so that granular
systems have been considered as genuine model systems for structure
formation on various length scales, including the formation of
planetesimals from interstellar dust and the formation of planets and
stars from accretion discs~\cite{astro}.

In most studies so far, models were inspired by dry granular systems,
where the dissipative contact interaction consists in the loss of a
certain fraction of the kinetic energy in every impact. Adding a small
amount of liquid to the granular system changes its properties
dramatically: while dry sand can flow freely similar to a liquid, wet
sand has properties of a plastic solid.  This difference in the
macroscopic behavior is reflected in a corresponding difference in
particle interactions \cite{HerminghausAdv2005}. The collisions of dry
granulates are typically purely \emph{repulsive} and characterized by
the coefficient of restitution $\varepsilon$ which specifies which
fraction of the kinetic energy is dissipated. Wet granular particles
are covered by a thin liquid film. When two particles come into
contact, the films merge and a capillary bridge is formed, exerting an
\emph{attractive} force on the particles. As the particles separate
from each other again, the bridge stays intact up to a critical
distance $d_\crit$. At this point the bridge ruptures
\cite{willet2000} and a fixed amount of energy is dissipated. Thus wet
granular particles are characterized by a hysteretic attractive
interaction and a {\it well defined energy} which is dissipated when a
capillary bridge ruptures.

The existence of a well defined energy scale (and corresponding
time scale), which is absent in dry materials, is the essential
microscopic ingredient not only of wet granulates but also of cohesive
gases. In fact the liquid bridge can be thought of as a particular
realisation of a more general cohesive force.
 A particularly important
aspect of free cooling in cohesive gases is the aggregation process
which sets in, when the kinetic energy falls below the bond breaking
energy. Wet granular systems may provide a realisation of various
aggregation models and so-called sticky gases~\cite{Carnevale90}, where particles move
diffusively or ballistically until they collide and get stuck to an
aggregate which is thereby growing. Such models have attracted a lot
of
interest
\cite{liang85,jiang93,jiang94,Carnevale90,alves06,dongen85,westbrook04,jullien84,trizac03,trizac95},
due to a wide range of applications ranging from the formation of dust
filaments, snowflakes and clouds to the size distribution and impact
probability of planetasimals in accretion discs.

Kinetic properties of granular gases have been discussed mainly for
dry materials. In particular, free cooling has been studied
extensively~\cite{BenNaim99,Nie02}, and it was shown that the
dissipative interactions are responsible for many novel phenomena,
unexpected from the kinetic theory of molecular gases: The particles'
velocities are not distributed according to a Maxwell- Boltzmann
distribution~\cite{Goldshtein95}, equipartition does not
hold~\cite{Huthmann97,Garzo99,Losert99}, a spatially homogeneous state
is generically unstable~\cite{Goldhirsch93}, and linear and angular
motion are correlated~\cite{Brilliantov07}.

Much less is known about {\it wet} granular media, which have been
addressed only recently
\cite{Thornton95,Lian98,Huang05,HerminghausAdv2005,Zaburdaev06,Fingerle06,Fingerle08,Fingerle2008b},
focussing on nonequilibrium phase transitions \cite{Fingerle2008b},
the equation of state~\cite{Fingerle08}, agglomeration
\cite{Ennis91,Thornton95,Lian98}, shear flow \cite{Huang05}, and
cooling in one dimension \cite{Zaburdaev06,Fingerle06}.

Structure formation in wet granulates during free cooling has
hardly been studied yet and is the focus of our paper which is
organized as follows. In Sec.~\ref{sec:model} we introduce the model
and discuss the decay of the average kinetic energy in
Sec.~\ref{sec:Cooling}. Aggregation is discussed in
Sec.~\ref{sec:percolation}, before we present conclusions in
Sec.~\ref{sec:conclusions}.  A short summary of our results has
appeared in \cite{WetPRL09}.

\section{Models\label{sec:model}}

In the present article, we are interested in the zero-gravity free
cooling dynamics of wet granular gases. We assume the particles to be
covered by a thin liquid film, as it is the case if the liquid
completely wets the particle material \cite{Israelachvili}. The particles
approach freely, until these surface films come into contact. The
liquid then rapidly accumulates around the contact due to the
interfacial forces. A capillary bridge forms at the contact, exerting
an attractive force on the grains due to its negative Laplace
pressure. This liquid bridge is stretched but stays intact (or even
continues to grow) as the particles move apart. The attractive force
thus remains until a certain critical separation $d_\crit$ is reached,
where the liquid neck becomes unstable and ruptures. As mentioned
above, the hysteretic formation and rupture of the bridge gives rise
to a characteristic loss of energy, $\Delta E$, which depends upon the
thickness of the liquid film wetting the grains.

\begin{figure}[h]
 \includegraphics[width=.45\textwidth]{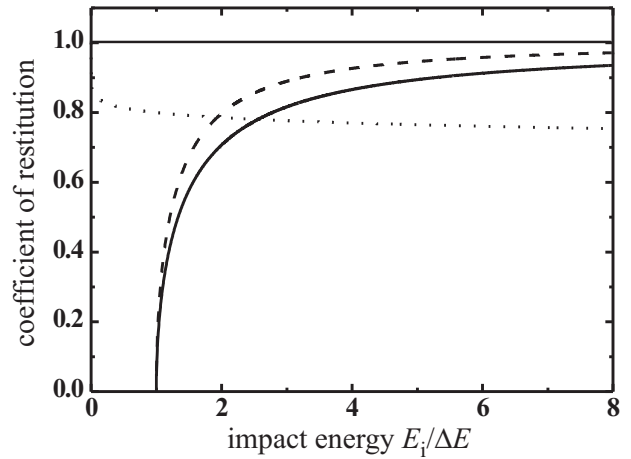}
 \centering
 \caption{Restitution coefficients for dry (dotted) and wet (solid and
   dashed) granular systems, plotted vs the impact energy in units of
   the wet energy loss, $\Delta E$.  The main feature in the wet case
   is the zero at $E_{\rm i} = \Delta E$, which is unchanged if the finite
   formation time for capillary bridges is taken into account (dashed
   curve).}
 \label{fig:Restitution}
\end{figure}

In order to design a suitable model, a few words on the details of
this process are in order. The formation of capillary bridges is
quite fast in real systems. Between typical grains of one millimeter
diameter it takes less than a millisecond. It is clear, however,
that this formation cannot in general be considered instantaneous if
the velocity of the impacting grains, $v_{\rm i}$, is large. If the
time scale of the impact process, which may be written as $d_\crit /
v_{\rm i}$, is of the same order or even smaller than the time of
capillary bridge formation, the accumulated liquid volume of the
bridge, and hence $\Delta E$, will be smaller than for slow impacts.
However, this will not greatly affect the main features of the wet
system, in particular as to its characteristic difference from the
dry granulate. In order to see that, we compare the effective
restitution coefficient of the dry and of the wet system. This is
shown in Fig.~\ref{fig:Restitution}, where the restitution
coefficient for the dry system is shown as the dotted curve. It
tends to be mildly depending on impact energy \cite{PoeschelBuch},
$E_{\rm i}$, with a negative slope throughout. The effective restitution
coefficient of the wet system, $\varepsilon_{\rm eff} =
\sqrt{1-\Delta E/E_{\rm i}}$, is shown as the solid curve, assuming
constant $\Delta E$. In strong contrast to the dry system, it has a
zero at $\Delta E/E_{\rm i} = 1$, and a markedly positive slope. This
illustrates the dramatic difference between these two systems. The
dashed line qualitatively accounts for the effect of finite
formation time of the capillary bridge. Since $\varepsilon_{\rm
eff}$ must stay below one, the difference between the solid and the
dashed curve is very limited, and the qualitative picture concerning
the comparison of dry and wet granular gases remains unchanged. 


Our system consists of $N$ identical and spherical particles with
diameter $d$ and mass $m$ in a three-dimensional cubic volume
$V=L^3$. The particles have a hard core interaction, such that two
particles are reflected elastically, if their centers of mass reach
the hard-core distance, which is the particle diameter $d$.

To account for the liquid film, a liquid bridge is allowed to form
between a pair of particles if they come close enough (``close
enough'' is specified later). When these particles are moving apart
and their distance exceeds the bond breaking distance $d_\crit$, the
liquid bridge will break and a fixed amount of kinetic energy $\Delta
E$ is dissipated; thereby, momentum is conserved and the relative
velocity $v_\text{rel}$ changes to $v_\text{rel}'$ according to
\begin{equation}
 \frac{\mu}{2} v_\text{rel}'^2  = \frac{\mu}{2} v_\text{rel}^2 - \Delta E
\end{equation}
with the reduced mass $\mu=m/2$. If, however, the relative kinetic
energy is smaller than $\Delta E$, the particles are elastically
reflected towards each other. The effect of the capillary force, which
is present in reality for distances up to $d_\crit$, is thus solely
modelled by the enrgy loss which occurs when $d=d_\crit$. This has
been shown before to be a very good
approximation~\cite{Fingerle2008b}, and enables event-driven
simulations as discussed below. For the formation of the liquid
bridge, we distinguish between two models:

In the \emph{thin film model}, the liquid bridge forms when the
particles touch, \emph{i.e.}~the distance of their centers is equal to
$d$. This model assumes that the liquid film covering the particles is
infinitesimally thin and the capillary bridges form a thin liquid
neck, which breaks off at the critical distance $d_\crit$.

\begin{figure}[h]
 \includegraphics[angle=270,width=.45\textwidth]{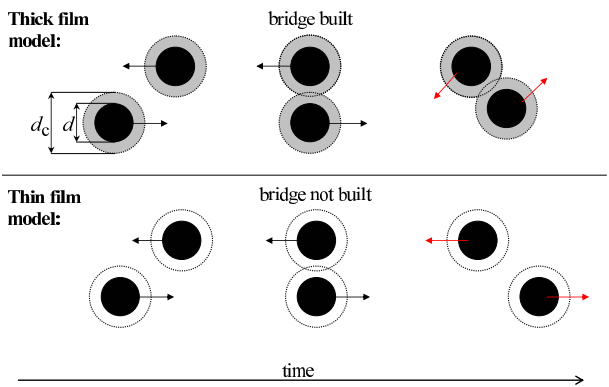}
 \centering
 \caption{Illustration of the thin film model and thick film model. In
   the thick film model, the liquid bridge forms, as soon as the bond
   breaking distances $d_\crit$ overlap. The same initial
   configuration in the thin film model does not create a liquid
   bridge, since the hard cores of the particles do not touch. Thus,
   the particles just pass by.}
 \label{fig:Models}
\end{figure}

In the \emph{thick film model}, a liquid bridge forms as soon as
particles come closer than the critical bond breaking distance
$d_\crit$. This model assumes that the outer diameter of the liquid
film is $d_\crit$ and its shape stays spherical and is not deformed by
the particles. Although this may seem unphysical, we include this case
in our study because similar assumptions have been used in many
simulation studies in earlier articles. As it will turn out, the
differences in most of the results are only minute. The two models are
illustrated in Fig.~\ref{fig:Models}.

In general there is some energy being transfered to the atomic degrees
of freedom of wet grains as well. In this paper we are going to
neglect this dissipation mechanism because it is usually small
as compared to the energy loss due to the breaking of
capillary bridges, especially if the granular temperature is small.  
However, we want to point out that such a dissipation mechanism can
easily be incorporated in the simulations, replacing the elastic
reflection by incomplete normal restitution. We restrict
ourselves here to perfectly smooth particles, such that translational and
rotational motion are decoupled. Furthermore, we investigate free cooling
only, so no gravity is present, and no energy is injected into
the system.

The particular way of accounting for the liquid film used in these
models makes it possible to use an event-driven simulation scheme.
The possible events are the reflection of the particles at the hard
core distance $d$ and the crossing of the bond-breaking distance
$d_\crit$. As mentioned above, we have previously compared event-driven
simulations of the wet system with full molecular dynamics simulations
integrating the equation of motion \cite{Fingerle2008b}. We found good
quantitative agreement in the results of both methods, justifying the
event-driven approach we chose exclusively for the present study.
 
We use dimensionless units such that $\Delta E = 1$, particle mass
$m = 1$ and particle diameter $d = 4$. The bond-breaking distance
is chosen as $d_\crit = 1.07 d$, unless noted otherwise, and 
volume fraction, $\phi=\pi d^3 /6 \cdot N/V$, is varied  from $\phi \approx 0.06\%$ up to
$15.6\%$.  We use periodic boundary conditions in the $x$- and $y$-direction
and hard walls in $z$-direction.

\section{Cooling Dynamics\label{sec:Cooling}}

We define the granular temperature $T = \frac{1}{3N} \sum_{i=1}^N m
{\bf v}_i^2$ and investigate its decay in time from a given initial
value $T_0\gg\Delta E$. In all our simulations we choose $T_0 = 45
\Delta E$. Simple arguments can be used to derive an analytical
form of the temperature decay. In each collsion a capillary bridge
ruptures with probability $P_{bb}$, giving rise to dissipation
of a fixed amount of energy, the bond-breaking
energy $\Delta E$. Particles collide with frequency $f_\text{coll}$,
so that the average loss of energy per unit time is given by:
\begin{equation}
 \frac{3}{2}\frac{dT}{dt} = - \frac{1}{2} \cdot f_\text{coll} \cdot \Delta
 E \cdot P_{bb}\, . 
\label{eq:dE/dt}
\end{equation}
The factor $\frac{1}{2}$ takes into account that two particles are
involved in one bond-rupture. 

\subsection{Early stage of cooling}

In the early stage of cooling the average kinetic energhy per particle
is much larger than the bond breaking energy, so that $P_{bb} \approx 1$
and almost every collision gives rise to dissipation by $\Delta
E$. For a dilute gas, the collision frequency
\begin{equation}
f_\text{coll} = 4 g(d) \sigma n \sqrt{\frac{T}{\pi m}} \label{eq:f_coll}
\end{equation}
is well established,  with the particle density
$n={N}/{V}$ and the pair correlation function at contact $g(d) =
\frac{(2-\phi)}{2(1-\phi)^3}$
(e.g.~\cite{PoeschelBuch}). 
The two models differ only 
in the cross section $\sigma$ (see Fig.~\ref{fig:Models}), which is
given by $\sigma = d\hc^2 \pi$ in the thin film model and $\sigma =
d_\crit^2 \pi$ in the thick film model.

The only temperature dependent quantity remaining on the right hand side of
Eq.~(\ref{eq:dE/dt}) is the collision frequency, $f_\text{coll} \sim
\sqrt{T}$ from (\ref{eq:f_coll}), giving rise to the following simple
equation: 
\begin{equation}
\frac{dT}{dt} \sim - \sqrt{T} \, ,
\end{equation}
which is solved by $T(t) \sim (t-t_0)^2$. Insertig the prefactors and
the initial value $T(0) = T_0$, one obtains, similar to Haff's
law \cite{HaffsLaw1983}, an analytical form of the decay of the temperature:
\begin{equation}
 T(t) = \left\{ \begin{array}{lll}
                 T_0\; (1-t/t_0)^2 \quad & \text{for } & t \leq t_0 \\
                 0                                  & \text{for } & t > t_0
                \end{array}
        \right. \label{eq:modelT(t)}
\end{equation}
with a charecteristic time scale
\begin{equation}
 t_0 = \frac{3 \sqrt{\pi m T_0}}{2 g(d) \sigma n \Delta E} \,. \label{eq:t0}
\end{equation}

\begin{figure}[h]
 \includegraphics[width=.49\textwidth]{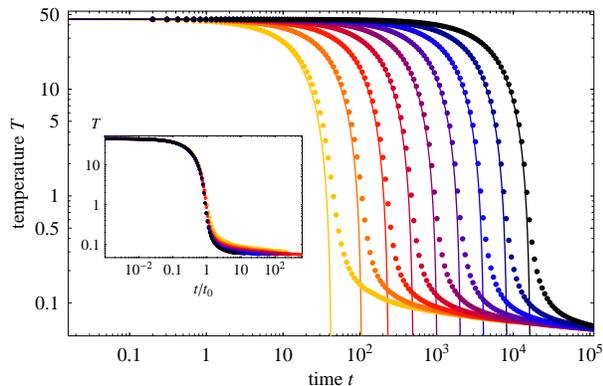}
 \centering
 \caption{(color online) Decay of the granular temperature $T$ for the
   thick film model and volume fractions (from left to right) $\phi =
   15.6\%, 7.81\%, 3.90\%,1.95\%, 0.98\%, 0.49\%, 0.24\%,0.12\%,$
   $0.061\%$. $N=262144$ particles are fixed. The corresponding solid
   lines show the analytic form (\ref{eq:modelT(t)}) with a decay to
   zero at time $t_0$, given in (\ref{eq:t0}). At that time
   the temperature of the \emph{simulated} granulate shows a rapid
   transition to a value below the bond-breaking energy \mbox{$\Delta
     E = 1$}.  In the inset temperature data are plotted versus scaled
   time $t/t_0$, such that data for different volume fractions
   collapse onto a single curve.}
 \label{fig:Temp}
\end{figure}

Note that, in this simplified model, the assumption that \emph{every}
collision causes an energy loss $\Delta E$ gives rise to a time-scale
$t_0$ after which all energy is dissipated. Even though this
assumption does not hold for all times in the simulation
(since the bonds do not break anymore if the relative kinetic energy
is too small), the timesacle $t_0$ has a clear physical relevance. 
It sets the time after which the temperature is
comparable to the bond-breaking energy $\Delta E$ and after which
persistent clusters will form. In Fig.~\ref{fig:Temp} the evolution of
the granular temperature $T$ from the simulation is compared to
(\ref{eq:modelT(t)}) for different volume fractions: $0.061\%
<\phi<15.6\%$. 

In the simplified cooling law (\ref{eq:modelT(t)}), the volume
fraction only enters into $t_0$. Hence we try to superimpose the data
by scaling time with $t_0$. As can be seen in the inset of
Fig.~\ref{fig:Temp}, the data obey the expected scaling well, except for the long time limit, which has different asymptotic behavior and is treated in the next section. 

\begin{figure}[h]
 \includegraphics[width=.49\textwidth]{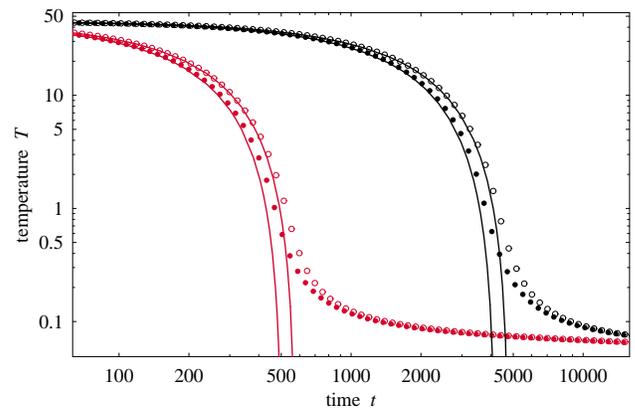}
 \centering
 \caption{(color online) Decay of the granular temperature $T$ for the
   thick film model ($\bullet$) and the thin film model ($\circ$) for
   volume fractions $\phi = 1.95\%$ and $\phi = 0.24\%$.}
 \label{fig:TempThickThin}
\end{figure}

In Fig.~\ref{fig:TempThickThin}, we compare data from the thin and
thick film model for two volume fractions, $\phi = 1.95\%$ and
$0.24\%$. The difference is solely due to different scattering
cross-sections, entering in $t_0$ (\ref{eq:t0}) and can be absorbed
into the rescaling of time by $t_0$.

\subsection{Late stage of cooling}
\label{sec:asymptotic_dynamics}

In the late stage of aggregation, when the system is strongly
aggregated, it becomes very unlikely that a capillary bridge ruptures.
Hence  we observe a very slow time evolution
of our system. The slow decrease of the temperature
can be understood with simple arguments.  The probability
$P_{bb}$ to break a bond is given by the probability to find a kinetic
energy larger than $\Delta E$:
\begin{equation}
P_{bb}=\int d^3v \,\theta(mv^2/2-\Delta E)\, w({\bf v}). 
\end{equation}
We approximate the velocity distribution $w({\bf v})$ by a Maxwellian
\begin{equation}
w({\bf v})=\biggl(\frac{m}{2\pi T(t)}\biggr)^{3/2} e^{-m v^2/(2T(t))} 
\end{equation}
and evaluate the above integral in the limit $T(t)/\Delta E \to
0$. The probability to break a bond becomes exponentially small in that
limit:
\begin{equation}
\label{eq:Pbb}
P_{bb}= \left(\frac{4\Delta E}{\pi T}\right)^{1/2}\, e^{-\Delta E/T}.
\end{equation}
The decrease of kinetic energy, as given by Eq.~(\ref{eq:dE/dt}),
is now dominated by the probability to break a bond. The collision
frequency $f_\text{coll}$ is not known for the clustered state, but is
expected to be proportional to $T^{1/2}$.
Using (\ref{eq:Pbb}) and $f_\text{coll}\propto T^{1/2}$ in the rate
equation (\ref{eq:dE/dt}) yields:
\begin{equation}
\label{eq:asymp_rate2}
\frac{dT}{dt} = - \, \gamma e^{-\Delta E/T}.
\end{equation}
The prefactor $\gamma$ is determined by the precise form of the
collision frequency. 
Separation of variables can be used to integrate Eq.(\ref{eq:asymp_rate2})
\begin{equation}
\label{eq:asymp_sol}
\int_{T_1/\Delta E}^{T/\Delta E} \, dx \,e^{1/x} =- \gamma(t-t_1)
\end{equation}
with the intital value $T_1=T(t_1)$.
In the asymptotic limit $T\to 0$ and $T_1\to 0$ with $T\ll T_1$, one
finds a logarithmically slow time decay of the temperature
\begin{equation}
\label{eq:T_assmyp_solution}
 \frac{T}{\Delta E} \sim \frac{1}{\ln (t)} \, 
\end{equation}
which is due to the very low probability to break
a bond, Eq.~(\ref{eq:Pbb}). This is in strong contrast to the
algebraic time decay observed for dry granular systems with
coefficient of restitution $\varepsilon < 1$. \cite{PoeschelBuch}

\begin{figure}[h]
 \includegraphics[width=.49\textwidth]{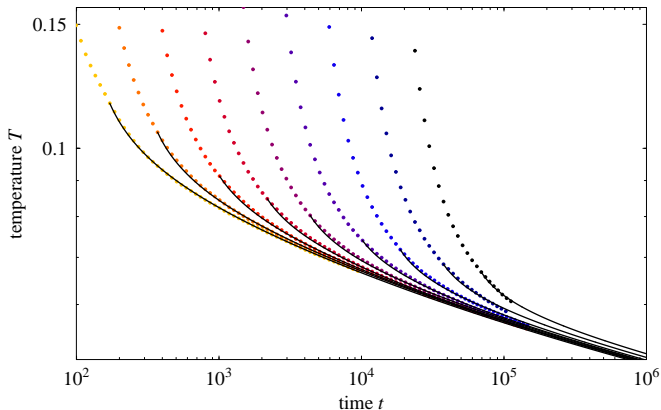}
  \centering
  \caption{(color online) Asymptotic time dependence for several
    volume fractions as in Fig.~\ref{fig:Temp}; data (dots) in
    comparison to the analytical results (lines)}
 \label{fig:asymptotic_time}
\end{figure}

In Fig.~\ref{fig:asymptotic_time},
the full solution (\ref{eq:asymp_sol}) is compared to the
simulation data, showing good
agreement. The unknown prefactor $\gamma$ is a fit parameter. It is
noteworthy that for all densities, the temperature seems to approach a
universal curve as $t \to \infty$.

\subsection{Partitioning of the energy into translational, rotational and internal degrees of freedom}

After the time $t_0$ has passed, stable clusters emerge. For the
definition of a cluster, we define particles as \emph{neighbors},
if a bridge is formed and the relative kinetic energy is not
sufficient to break it. This makes sure that particles which are just
``passing by'', are not considered neighbors. A \emph{cluster} is a
set of particles connected through this neighbor-relationship. Hereby
we refer to the cluster mass $m$ as the number of particles a cluster
contains. Clusters defined in this way are not truly stable. Particles
belonging to the cluster are occasionally kicked out, if hit by a
very energetic particle.

For a more detailed understanding of the system, we investigate the
cooling dynamics on the cluster level, and determine how energy is
partitioned among the degrees of freedom. We split the total temperature $T$
into three constituents, the \emph{translational} temperature defined
via the center-of-mass velocities of the clusters, the \emph{rotational}
termperature defined via the angular momenta of the clusters, and the
\emph{internal} temperature describing the relative movement of the
particles inside a cluster. These three temperatures are defined as
follows. 

Our definition of neighborhood relations gives rise to $n_\cl$
distinct clusters numbered by $i = 1,...,n_\cl$. We denote by
$\mathcal{N}_i$ the $i$-th cluster with $m_i$ particles.  Its centre
of mass position and velocity are given by:
\begin{equation}
 {\bf R}_i =\frac{1}{m_i} \sum_{\nu \in \mathcal{N}_i} {\bf r}_{\nu}
\quad \mbox{and} \quad {\bf V}_i =\frac{1}{m_i} \sum_{\nu \in \mathcal{N}_i} {\bf v}_{\nu} \,.
\end{equation}
Note that single particles with $m_i = 1$ are also considered as clusters.

The center of mass movement of each cluster has $f_{\text{trans},i} =
3$ translational degrees of freedom, so that the total number of
translational degrees of freedom of these clusters is simply $3
n_\cl$. Homogeneous cluster translations are thus characterized by 
the \emph{translational} temperature
\begin{equation}
  T_\text{trans} := \frac{2}{3 n_\cl} 
\cdot \sum_{i=1}^{n_\cl} \frac{m_i}{2}{\bf V}_i^2  \, .
\end{equation}

Analogously, the \emph{rotational} temperature describes the energy in
homogeneous cluster rotations. The angular momentum, ${\bf L}_i$, of
cluster $i$ is given in terms of the relative particle positions
$\tilde{\bf r}_{i,\nu} = {\bf R}_i - {\bf r}_\nu$ and velocities
$\tilde{\bf v}_{i,\nu} = {\bf V}_i - {\bf v}_\nu$
\begin{equation}
 {\bf L}_i = \sum_{\nu \in \mathcal{N}_i}\tilde{\bf r}_{i,\nu} 
\times \tilde{\bf v}_{i,\nu} \, .
\end{equation}
The rotational energy of cluster $\mathcal{N}_i$ with $m_i>2$ is thus
given by 
\begin{equation}
 E_{\text{rot},i} = \frac{1}{2} {\bf L}_i \, \b{\b{I}}_i^{-1} \,  {\bf L}_i \, ,
\end{equation}
where the moment of inertia tensor $\b{\b{I}}_i$ is defined in the
usual way. The case $m_i = 2$, requires special treatment, since the inertia
tensor is singular. The rotational energy of a dimer can be easily
calculted to $E_{\text{rot},i}=({\bf v}_1-{\bf v}_2)_{\perp}^2/4$, where $
({\bf v}_{1}-{\bf v}_{2})_{\perp}$ denotes the relative velocity
perpendicular to the axis of the dimer.
The \emph{rotational} temperature is thus
\begin{equation}
 T_\text{rot} := \frac{2}{\sum_{i=1}^{n_\cl} f_{\text{rot},i}} 
\cdot  \sum_{i=1}^{n_\cl} E_{\text{rot},i} \, , \label{eq:Trot}
\end{equation}
with $f_{\text{rot},i} = 2$ for dimers and $f_{\text{rot},i} = 3$ for
larger clusters.

All the left-over kinetic energy $E_\text{int}$ describes the relative
movement of particles inside a cluster and contributes to the
\emph{internal} temperature. Each cluster has a total of $3 m_i$
degrees of freedom, so that the remaining number for internal degrees
of freedom is $f_{\text{int},i} = 3m_i - f_{\text{trans},i} -
f_{\text{rot},i}$.
The internal temperature $T_\text{int}$ is:
\begin{equation}
 T_\text{int} := \frac{2}{\sum_{i=1}^{n_\cl} f_{\text{int},i}} \cdot  \sum_{i=1}^{n_\cl} E_{\text{int},i} \, .\label{eq:Tint}
\end{equation}

\begin{figure}[t]
 \includegraphics[width=.49\textwidth]{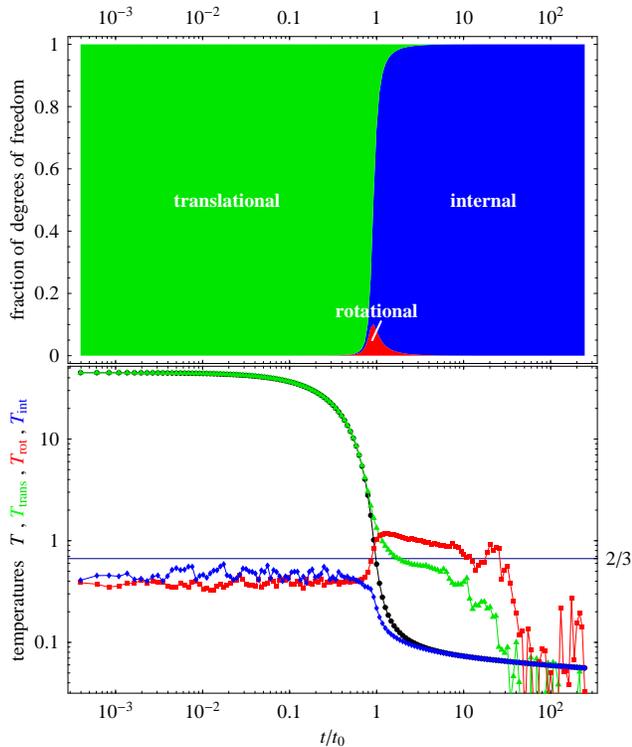}
 \centering
 \caption{(color online) Top: Division of the total $3N$ degrees of
   freedom into the \emph{translational}, \emph{rotational} and
   \emph{internal} parts, dependent on time.  Bottom: Evolution of the
   \emph{total} ($\bullet$, black), \emph{translational}
   ($\blacktriangle$, green),
   \emph{rotational} ($\scriptstyle\blacksquare$, red), and \emph{internal}
   ($\blacklozenge$, blue) granular temperatures. Data for $N=262144$
   particles and volume fraction $\phi = 1.95\%$ are shown; the
   behavior is qualitatively the same for all investigated system
   sizes. The horizontal line at $2/3$ corresponds to the bond
   breaking energy.}
 \label{fig:DegOfFreedom}
\end{figure}

Fig.~\ref{fig:DegOfFreedom} (top) shows how the total of $3N$ degrees
of freedom divide up into translational, \emph{rotational}, and
\emph{internal} degrees of freedom. The corresponding temperatures are
shown in the lower half of the figure. As one might expect, for $t \ll
t_0$ almost all degrees of freedom are \emph{translational}, since
most clusters are just single
particles, 
and $T_\text{trans} \approx T$. Keeping in mind that two particles are
only defined as neighbors if their relative velocity is not
sufficient to break the bond, only \emph{stable} clusters (mostly
dimers) enter the \emph{internal} and \emph{rotational} temperatures,
and therefore $T_\text{rot}, T_\text{int} < \frac{2}{3}\Delta E =
\frac{2}{3}$ for $t/t_0<1$. \footnote{The factor $\frac{2}{3}$ is due
  to the relation $\frac{2}{3} \bar{E}_\text{kin} = T$ between
  temperature and energy.}

In the transitional regime $t \approx t_0$, when the number of
intermediate size clusters increases, the rotational degrees of
freedom become important. Larger objects can have higher rotational
energies without rupture \footnote{roughly speaking, the maximum
  rotational energy $E_\text{rot,max}$ of a cluster with radius $r$
  and mass $M$ is $E_\text{rot,max} \sim M r^2 \omega^2$, where the
  maximum rotational frequency $\omega$ is limited by the centrifugal
  force $F \sim M \omega^2 r$. This yields $E_\text{rot,max} \sim
  r$. In our case the bond breaking energy $\Delta E$ is related to
  the maximum force $F$ on the particles by $\Delta E \sim F \cdot
  (d_\crit-d\hc)$, with the freely movable distance of a particle
  $(d_\crit-d\hc)$.}, therefore the growing clusters obtain energy
from caught particles, and thus $T_\text{rot}$ increases until
reaching the value of $T_\text{trans}$. After that, the energy of the
incoming lumps is not sufficient to increase $T_\text{rot}$ any
further.

In contrast to the homogeneous cluster \emph{rotations}, the
\emph{internal} degrees of freedom which have higher energies than
$\Delta E$ will in most cases result in a bond rupture, independent of
the cluster size. Therefore, $T_\text{int}$ decreases monotonically. At
late times $t \gg t_0$, large clusters have formed, thus almost all
degrees of freedom are \emph{internal} and $T \approx T_\text{int}$.

\section{Aggregation\label{sec:percolation}}

When the average kinetic energy per particle is comparable to the bond
breaking energy, $t\sim t_0$, the system starts to form aggregates,
which seem to grow in a self-similar process. In the following we are
going to analyze these aggregates and compare them to cluster-cluster aggregation \cite{Jullien1987} models. As time proceeds,
larger and larger clusters are formed. We observe a
spanning or percolating cluster for all finite densities, and
ultimately all particles and clusters have merged into a single
cluster.

\begin{figure}[t]
\includegraphics[width=.4998\textwidth]{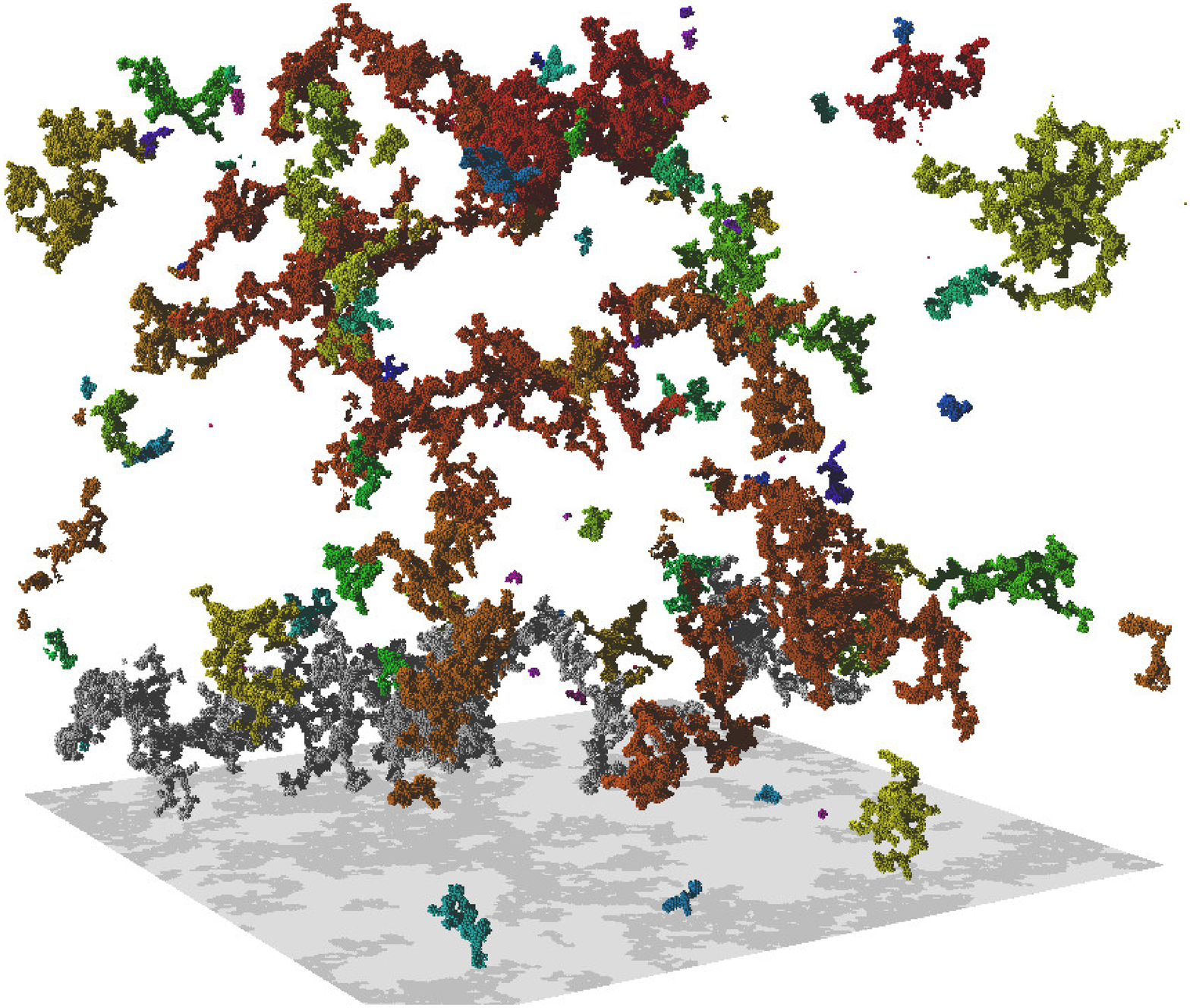}
\includegraphics[bb=600 530 620 531,width=.01\textwidth]{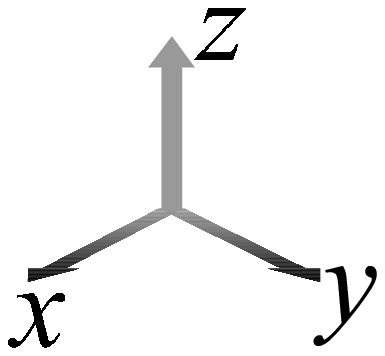}
 \centering
 \caption{(color online) Snapshot of the system with volume fraction
   $\phi = 0.48\%$ and $N=262144$ particles taken at time $t \approx
   12 t_0$; the largest cluster (grey) contains 22\% of the
   particles. Particles of the same cluster have the same color shade.}
 \label{fig:SystemSnapshot1}
\end{figure}

\begin{figure}[t]
\includegraphics[width=.4998\textwidth]{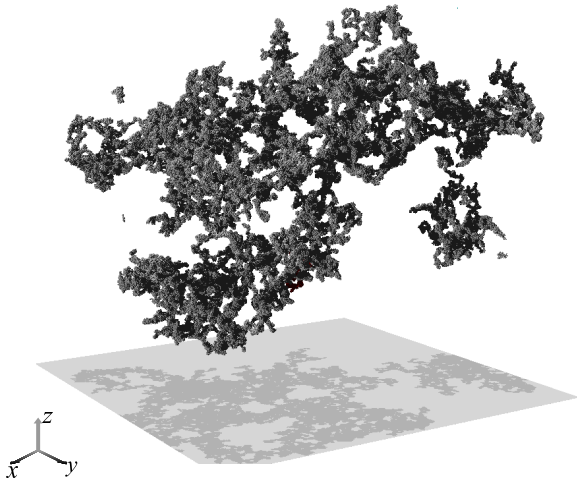}
\includegraphics[bb=600 530 620 531,width=.01\textwidth]{CoordinateSystem.eps}
 \centering
 \caption{(color online) Same as Fig.~\ref{fig:SystemSnapshot1} for $t
   \approx 52 t_0$; the largest cluster contains 99\% of the
   particles.}
 \label{fig:SystemSnapshot2}
\end{figure}

Figs.~\ref{fig:SystemSnapshot1} and \ref{fig:SystemSnapshot2} 
show snapshots of a system at $t= 12 t_0$ and $t= 52 t_0$ with small
volume fraction, $\phi=0.48\%$.  At the smaller time the system is not
yet percolating, even though rather large clusters have already
formed, the largest one (in grey) contains $22\% $ of all particles. The
second snapshot, taken at a much longer time, shows a spanning
cluster. At such large times the average kinetic energy is much
smaller than the bond breaking energy ($T \approx 0.06 \Delta E$), so that
bonds almost never break up. The cluster shown is already well beyond
the critical time for percolation with $99\%$ of the particles in the
cluster.

\begin{figure}[h]
\begin{minipage}[h]{.39\textwidth}
\includegraphics[width=\textwidth]{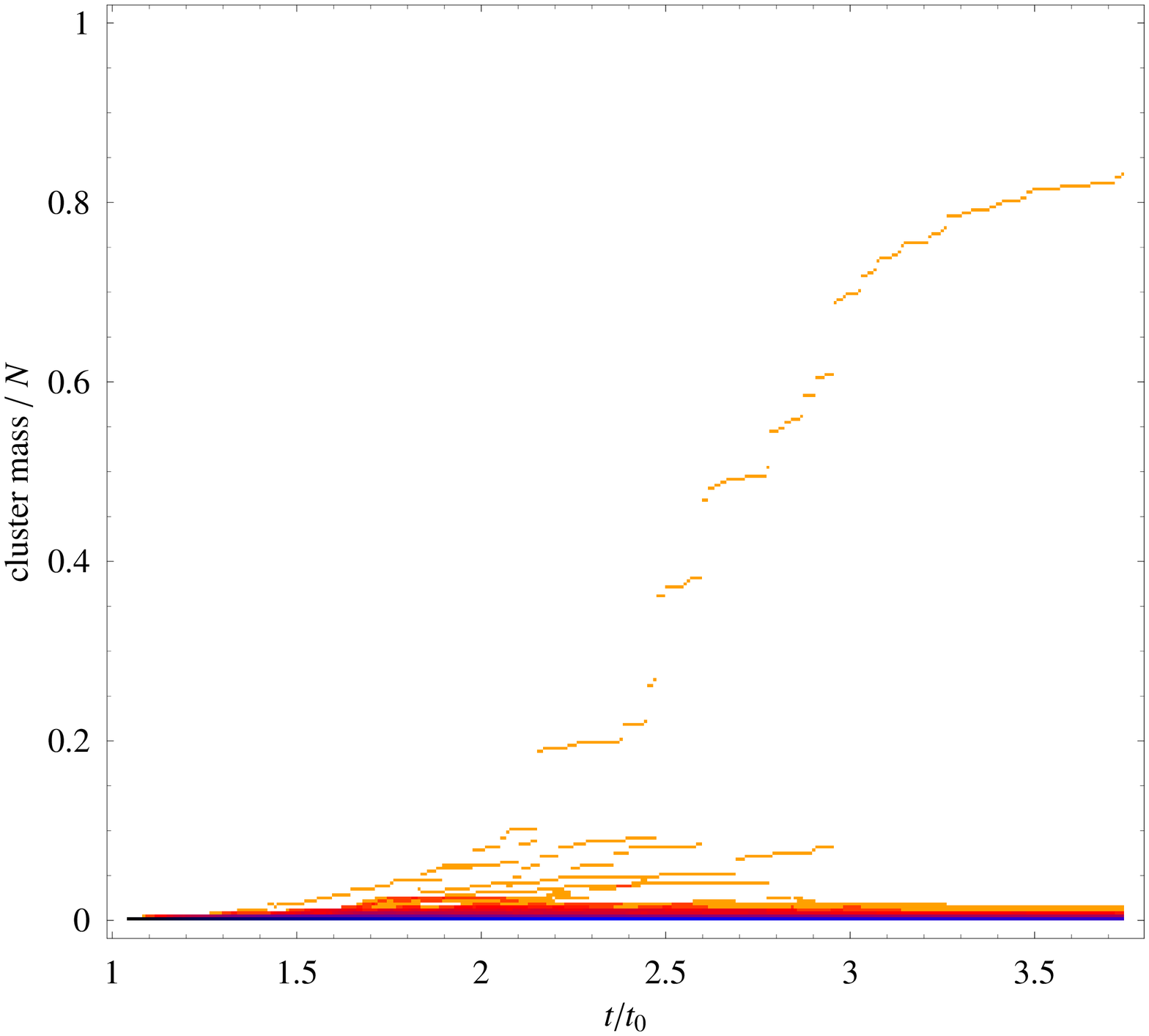}
\end{minipage}
\begin{minipage}[h]{.08\textwidth}
 \includegraphics[height=5cm]{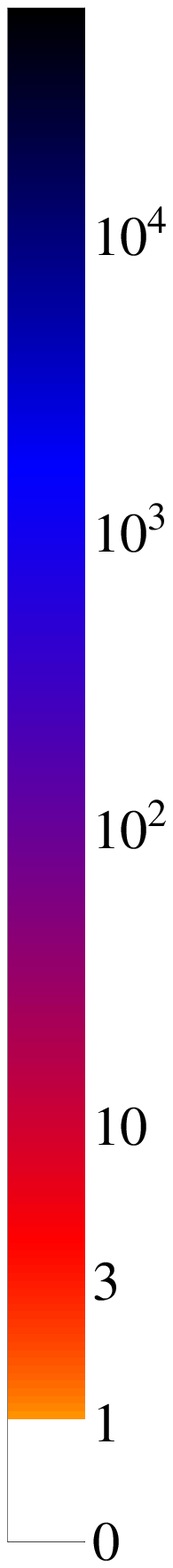}
\end{minipage}
\centering
 \caption{Histogram of the cluster mass distribution dependent on
   time, for volume fraction $\phi = 3.9\%$ and $N=262144$. The number
   of clusters at the respective time and size is color coded on a
   logarithmic scale so that the single largest cluster is visible. At
   $t \approx 2.5 t_0$ one can see the large cluster emerging, clearly
   distinguishable from the rest of the distribution.}
 \label{fig:ClusterSizeDist}
\end{figure}

Fig.~\ref{fig:ClusterSizeDist} shows the evolution of the cluster mass
distribution $N_m(t)$, which is the number of clusters containing $m$
particles at time $t$. One can clearly see that after some time, $t
\approx 2.5 t_0$, which depends on volume fraction, the largest
cluster emerges from the rest of the distribution. For all volume
fractions a gelation transition was observed at the percolation time
$t_c>t_0$. The critical behavior of the gelation transition is still
controversial. Since aggregation is a {\it nonequilibrium} process,
there is a priori no reason that it should be in the same universality
class as the corresponding equilibrium percolation transition. Yet
there is some evidence in favour of this conjecture.  Gimel \emph{et
  al.}~\cite{Gimel1995} observe a crossover from self-similar growth
at small times and volume fractions -- called the flocculation regime
-- to the percolation regime around $t_c$. In the latter
they observe critical exponents as in standard percolation
theory. Kolb and Herrmann \cite{Kolb1985} on the other hand obtain
values for the fractal dimension of the percolating cluster, distinct
from percolation theory as well as from flocculation theory. Both
studies refer to diffusion limited cluster-cluster aggregation.

In this paper we do not analyze the gelation transition in detail but
defer such a discussion to future work.
Instead we investigate two regimes in detail in the following:

a) The self-similar growth process, or flocculation regime, which is
present for small times and volume fractions. 

b) The properties of the final cluster which emerges, when (almost) all
particles have aggregated to form one large cluster.

\subsection{Self-similar growth\label{sec:SelfSimilarGrowth}}

\subsubsection{Fractal dimension of the aggregates\label{fractal_aggregates}}

A central quantity of aggregation models is the fractal dimension of
the aggregates. It is usually determined from the radius of gyration
as a function of cluster mass. We consider a cluster of $m$ particles with
positions $({\bf r}_1, ... , {\bf r}_m)$ and define its radius of
gyration by (see e.g.~\cite{OnGrowthAndForm})
\begin{equation}
 r_\text{g}^2(m) = {\frac{1}{m} \sum_{i=1}^m ({\bf r}_i - \bar{\bf r})^2 } \text{\quad with \quad} \bar{\bf r} = \frac1m \sum_{i=1}^m {\bf r}_i \, .
\end{equation}
If the clusters are fractal we expect a  scaling relation for large
$m$ of the form
\begin{equation}
\label{scaling_r_gyr}
 r_\text{g} \sim m^{1/D_\text{f}}
\end{equation}
which yields the fractal dimension $D_\text{f}$. This method is
commonly used in aggregation models, where particles move diffusively,
ballistically, or are interacting and stick to the aggregate once they
touch it \cite{alves06,westbrook04,jullien84,meakin91}. 

\begin{figure}[h]
 \includegraphics[width=.49\textwidth]{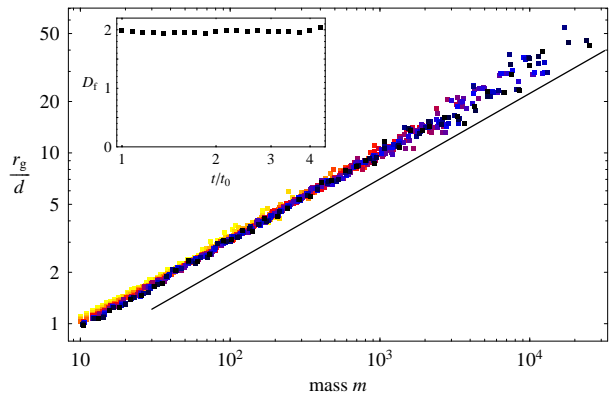}
 \centering
 \caption{(color online) Radius of gyration as a function of cluster
   size for a system of 262144 particles at volume fraction
   $\phi=1.96\%$; different colors/shades correspond to simualtion times
   between $t_0$ (yellow/light gray) and $4t_0<t_c$ (black)); The slope of the solid line corresponds to $D_\text{f} = 2$; inset: fractal
   dimension as a function of time, extracted from the slope of the
   curves in the main figure.}
 \label{radius_gyration}
\end{figure}

In Fig.~\ref{radius_gyration} we show the radius of gyration for a
sytem of $262144$ particles at volume fraction $\phi=1.96\%$. Several
snapshots of the ensemble of growing clusters have been taken at times
$t_0<t<t_c$ with the percolation time $t_c$, when a spanning cluster
is first observed. The data scale well according to 
Eq.(\ref{scaling_r_gyr}), some scatter is observed for the largest masses,
corresponding to times close to the percolation transition.

In contrast to aggregation models, where the clusters are static and
do not break up, we occasionally do observe the breaking of bonds. In
addition there are internal deformations of the clusters during
growth, so that the fractal dimension could depend on time. We have
therefore checked the relation between $m$ and $r_\text{g}(m)$ for
many instances of time and show the fractal dimension as a function of
time in the inset of Fig.~\ref{radius_gyration}. As can be seen from
the Figure, there is no systematic dependence on time, and the fractal
dimension is close to $D_f=2$.

\subsubsection{Cluster size distribution\label{sec:ClusterScaling}}

\begin{figure}[h]
 \includegraphics[width=.49\textwidth]{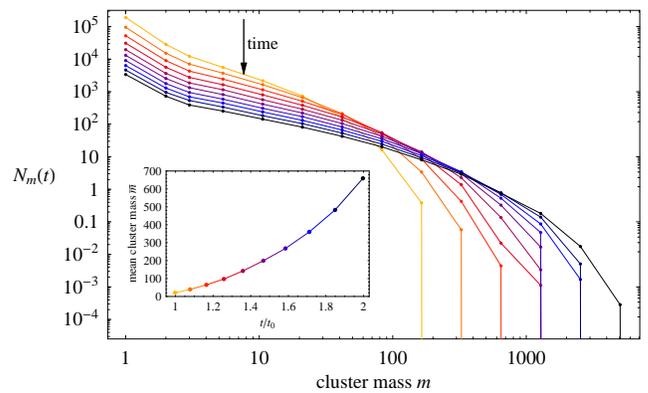}
 \centering
 \caption{(color online) The cluster mass distribution $N_m(t)$. The
   different graphs represent different times, which are
   increasing from top to bottom (left side of the graph). The inset
   shows how the mean cluster mass increases during the investigated
   time period.}
 \label{fig:ClusterSizeNoScaling}
\end{figure}

All information about the connectivity of the clusters is contained in
the cluster size distribution $N_m(t)$, the number of clusters of size
$m$ at time $t$. In Fig.~\ref{fig:ClusterSizeNoScaling} we show
$N_m(t)$ for a system with $\phi = 1.96 \%$ and $N=1048576$. The time
interval has been chosen such that $t_0 < t < 2t_0 < t_c \approx 4 t_0$ (for
this volume fraction). In this time interval the mean cluster mass increases roughly by a factor of 30. 

It has been suggested (\emph{e.g.}~\cite{meakin91}) that for aggregating
systems the mass distribution evolves towards a self-preserving
scaling form, independent of the initial distribution:
\begin{equation}
 N_m(t) = m^{-\theta} f\big(m / \bar{m}(t)\big) \, ,\label{eq:ClusterScaling}
\end{equation}
where the time dependence is only contained in the mean cluster mass
\begin{equation}
\bar{m}(t)=\frac{\sum_{m=1}^{\infty} m^2 N_m(t)}{\sum_{m=1}^{\infty} m N_m(t)}\, . \label{Eq:mbarDef}
\end{equation}
This scaling form has been applied sucessfully to various aggregating
systems
\cite{vicsek84,botet84,jiang93,jiang94,dongen85,meakin91,trizac95},
involving fractal as well as non-fractal objects. Mass conservation
requires $\theta = 2$ \cite{meakin91}. 

\begin{figure}[h]
 \includegraphics[width=.49\textwidth]{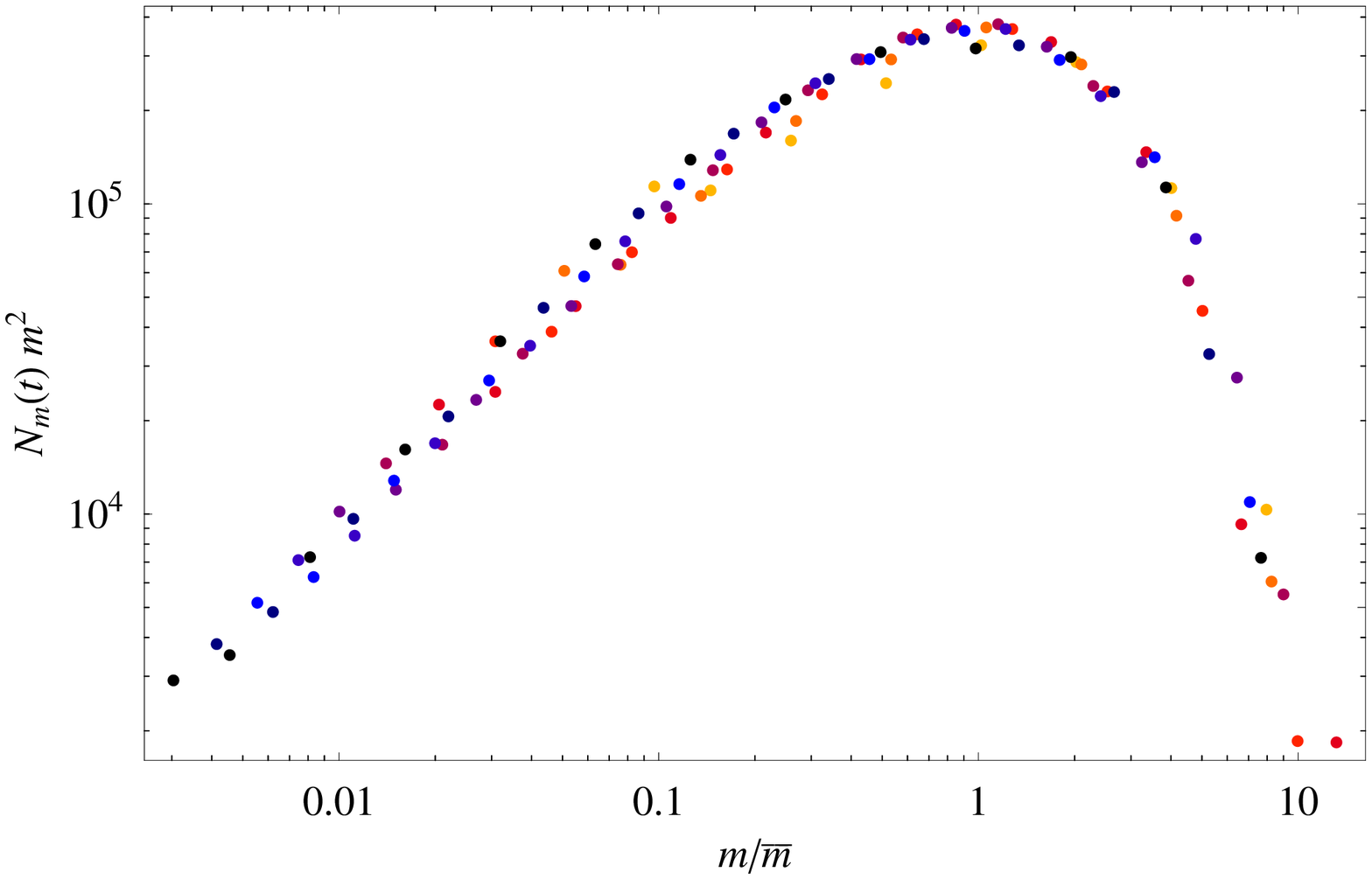}
 \centering
 \caption{(color online) Rescaled cluster size distribution $f(m /
   \bar{m}) = N_m(t) \cdot m^{2}$ from eq.~(\ref{eq:ClusterScaling})
   versus the normalized cluster mass $m / \bar{m}$. The color coding
   as in Fig.~\ref{fig:ClusterSizeNoScaling} is used.}
 \label{fig:ClusterSizeScaling}
\end{figure}

We plot in Fig.~\ref{fig:ClusterSizeScaling} the scaling function $f(m / \bar{m})= N_m(t) m^2 $ for the same data sets as in 
Fig.~\ref {fig:ClusterSizeNoScaling}. We expect scaling to hold only in the
aggregation regime, \emph{i.e.}~for times not too close to $t_c$, where the
system gels (see sec.~\ref{sec:meanClusterMass}). Hence we restrict ourselves in Fig.~\ref{fig:ClusterSizeScaling}
to times $t_0 < t < 2t_0$. We have also left out the data
points for $m=1$, \emph{i.e.}~clusters consisting of single particles. As can
be seen from Fig.~\ref{fig:ClusterSizeScaling} the data scale very
well. Deviations occur only for times close to the percolation
transition (not shown here), where they should be expected.

\subsubsection{Number of clusters \label{sec:meanClusterMass}}

Another characteristic of a realisation of clusters is simply the
total number of clusters $n_\cl(t) = \sum_{m=1}^{\infty} N_m(t)$,
which decreases as aggregation proceeds. As
long as the system is in the scaling regime (\emph{i.e.}~relation
(\ref{eq:ClusterScaling}) is fullfilled), the mean cluster mass,
$\bar{m}(t)$ and the number of clusters are simply related:
$\bar{m}(t) \sim n_\cl^{-1}$. However, as mentioned above, the scaling relation
(\ref{eq:ClusterScaling}) only holds in the aggregation regime and is
expected to break down as the percolation transition is approached.
At that point, $\bar{m}$ should diverge due to the formation of a
spanning cluster. 
On the other hand, there is
still a large number of smaller clusters coexisting with the
macroscopic cluster, so that $n_\cl / N$ remains finite at the
percolation transition. 

The aggregation of particles to larger objects has been investigated
for various \emph{ballistic aggregation} models
\cite{Carnevale90,jiang94,trizac03,family85}, where spherical
particles of mass $m=1$ and diameter $d=d_0$ move ballistically, until
two of them collide to form clusters irreversibly. In a particularly
simple model, one assumes that two colliding particles form one larger
spherical particle with conserved momentum and a mass $m$ equal to the
sum of the two particles masses, so that $m$ is always equal to the
number of \emph{initial particles} contained in a given cluster. For
spatial dimension $D$, the diameter increases like $d=m^{1/D} d_0$,
assuming the particles to be compact spheres which conserve volume
when merging. For this model, a mean field theory \cite{jiang94} and
simple scaling arguments \cite{Carnevale90,trizac03} yield the
dependence of the expected average mass $\bar{m}$ on time like
$\bar{m} \sim t^\xi$ with an exponent $\xi = 2D/(D+2)$ (assuming
$t_0=0$).

Since the aggregating clusters in our system are not compact, but
fractal objects with fractal dimension $D_\text{f}$, the assumption
for the diameter $d \sim m^{1/D}$ does not hold and must be changed to
$d \sim m^{1/D_\text{f}}$. With this assumption, we follow the scaling
arguments of Trizac \emph{et al.} \cite{trizac03}, and find the
scaling relation between $\bar{m}$ and $t$.

\begin{figure}[h]
 \includegraphics[width=.49\textwidth]{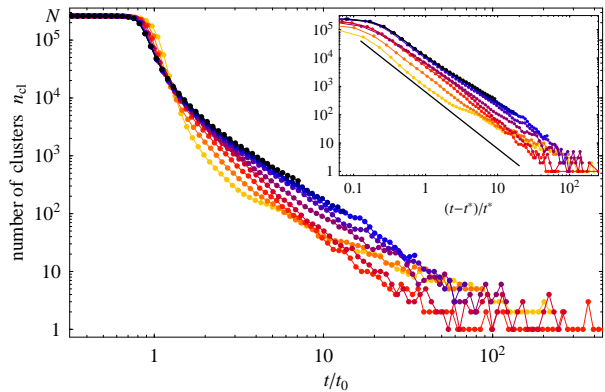}
 \centering
 \caption{(color online) Evolution of the mean cluster mass. Labeling
   and parameters as in Fig.~\ref{fig:Temp}. For the inset, the origin
   of the time-axis has been shifted to the transition point
   $t^*$ to investigate the scaling relation $n_\cl \sim (t - t^*)^{-\xi}$.
   The solid line has a slope of $-2$.}
 \label{fig:MeanClusterSize}
\end{figure}

We assume that the number of clusters per volume, $n_\cl$, is reduced
by one whenever two clusters collide:
\begin{equation}
 dn_\cl / dt \sim - f_\text{coll} \cdot n_\cl \, .
\end{equation}
The collision frequency \cite{PoeschelBuch}
is approximately given by $f_\text{coll} \sim d^{D-1} n_\cl {v}$
with $d \propto r_\text{g}$ the linear dimension of the cluster and ${v}$
its typical velocity. The average momentum should
scale as $p \sim m^{1/2}$ \cite{trizac03}, and therefore
\begin{equation}
 {v} = p/m \sim m^{-1/2} \sim n_\cl^{1/2} \, .
\end{equation}
Plugging in all these scaling relations as well as ${m} \sim
r_\text{g}^{D_\text{f}}$, one obtains:
\begin{equation}
  \frac{dn_\cl}{dt} \sim -n_\cl^2 \cdot {v} \cdot {d}^{D-1} 
  \sim - n_\cl^{5/2 - (D-1)/D_\text{f}} \, ,
\end{equation}
which is solved by
\begin{equation}
 n_\cl \sim (t-t^*)^{-{2D_\text{f}}/({3D_\text{f} - 2D + 2})} \, , \label{eq:nclScaling}
\end{equation}
where the integration constant $t^*$ is the onset of cluster growth. In
our context $t^* \approx t_0$ \footnote{As one can see in the main
  plot of Fig.~\ref{fig:MeanClusterSize}, the actual onset of cluster
  growth is not exactly at $t_0$, but a little bit earlier.}. This implies the following growth law for the mean cluster mass in the scaling regime:
\begin{equation}
 \bar{m} \sim (t-t^*)^\xi \text{\quad with \quad} \xi = \frac{2 D_\text{f}}{3 D_\text{f} - 2 D + 2} \, , \label{eq:MassScaling}
\end{equation}
which generalises the result for compact objects, $\xi = 2D/(D+2)$ with $D = D_\text{f}$ to fractal ones with $D\neq D_\text{f}$.

In Fig.~\ref{fig:MeanClusterSize} we show how the number of clusters decreases over time as larger and larger aggregates form for $t>t_0$.

The inset of fig.~\ref{fig:MeanClusterSize} investigates the scaling behavior (\ref{eq:nclScaling}), with the origin of the time axis shifted to the transition point $t^*$. One can see that the slope of $\xi = 2$, obtained from (\ref{eq:MassScaling}) for $D = 3$ and $D_\text{f} = 2$ is in good agreement with the simulation.

\subsection{Properties of the asymptotic cluster\label{sec:LargestCluster}}

The fractal dimension of the largest cluster -- well beyond the
percolation transition for most volume fractions -- will be the main focus of
this section. In particular we determine its fractal dimensions and
coordination numbers.

\subsubsection{Fractal dimension from radius of gyration\label{sec:LCrgFractalDim}}

One way to determine the fractal dimension is the radius of gyration,
as was done in Sec.~\ref{fractal_aggregates} for aggregates. Here,
however, we only have one large cluster and have to find a way to
obtain the function $r_\text{g}(m)$ as a function of cluster size
$m$. We implement this in following way: Starting from a random
particle of the cluster, we mark all particles that can be reached
through $i$ neighbor-to-neighbor steps. Thus, for every $i$, we get a
\emph{partial cluster} with $m(i)$ particles and radius of gyration
$r_\text{g}(i)$, which yields the scaling relation $r_\text{g} \sim
m^{1/D_\text{f}}$ and the fractal dimension $D_\text{f}$. For good
statistics, we repeat this procedure 100 times (each with a different
initial particle) and average over the obtained values of
$r_\text{g}$. Note furthermore that the procedure takes care that no
particle is marked a second time, in order to make sure that one does
not go through the cluster several times because of the periodic
boundary conditions.

\begin{figure}[t]
 \includegraphics[width=.49\textwidth]{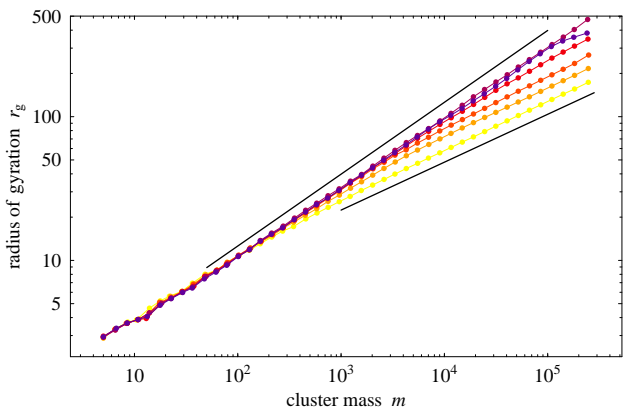}
  \centering
  \caption{(color online) Radius of gyration dependent on the mass of
    the partial cluster at simulation time $t \approx 27 t_0$. The
    particle number is fixed $N = 262144$ and the volume fractions are
    (from bottom to top) $\phi = 15.6\%, 7.81\%, 3.90\%, 1.95\%,
    0.98\%, 0.49\%$. The lines along the data points are the
    respective fits. The outer solid lines have slopes $1/2$ (top)
    and $1/3$ (bottom) corresponding to fractal dimensions of $2$ and
    $3$, respectively.}
 \label{fig:rgDimScaling}
\end{figure}

In Fig.~\ref{fig:rgDimScaling} we show the results of this procedure
for the radius of gyration $r_\text{g}$ as a function of $m$ for
different densities.  For high volume fractions we are well beyond the
percolation transition and hence expect $D_{\text{f}}=3$ on the
{\it largest} length scales of the cluster. This is clearly seen in
Fig.~\ref{fig:rgDimScaling}, e.g. for $\phi = 15.6\%$ and $10^3 < m
<10^5$. On smaller length scales, however,
we find a fractal dimension $D_{\text{f}} \approx 2$. For smaller volume
fractions, the crossover to $D_{\text{f}}=3$ happens at larger masses
and hence the ``interior'' region extends to larger scales.


\subsubsection{Fractal dimension from box counting algorithm\label{sec:LCFractalDim}}

To further investigate the Hausdorff dimension of the largest cluster at
{\it intermediate} length scales, we use the
box counting algorithm \cite{grassberger83,hentschel83}. The system is
divided into sub-boxes of edge length $L_\bx$. Then each box which
contains or hits at least one particle is marked. In this way, we find
the number of boxes $N_\bx$ necessary to cover the whole
cluster. 
This number should scale with $L_\bx$ like
\begin{equation}
 N_\bx \sim L_\bx^{-D_\text{f}} \, ,
\end{equation}
with the Hausdorff dimension $D_\text{f}$.

On length scales much smaller than the particle diameter, $L_\bx \ll
d\hc$, the system obviously behaves three-dimensionally. In this
regime, the number of filled boxes $N_\bx$ is just the volume fraction
$\phi$ times the total number of boxes $N_\text{box,tot} =
L^3/L_\bx^3$, therefore:
\begin{eqnarray}
N_\bx = \frac{\phi L^3}{L_\bx^3} \, . \label{eq:fracDimSmallL}
\end{eqnarray}

Since our system is finite and contains a system-spanning cluster, the
scaling behavior on large length scales $L_\bx \approx L$ should also
be three dimensional. On this length scale, almost all the boxes should
be filled, so that
\begin{eqnarray}
N_\bx = \frac{L^3}{L_\bx^3} \, . \label{eq:fracDimLargeL}
\end{eqnarray}
In particlar, the relation must include the point $(L_\bx, N_\bx) =
(L,1)$, since a box of the system size includes all particles and will
certainly be marked.

\begin{figure}[h]
 \includegraphics[width=.45\textwidth]{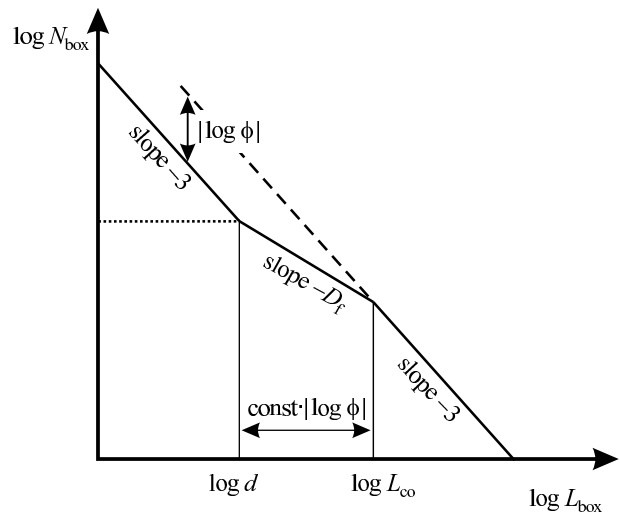}
 \centering
 \caption{Schematic double logarithmic plot of the box size $L_\bx$ versus the number of boxes $N_\bx$ of that size needed to cover the cluster. The negative slope is the fractal dimension. We expect three scaling regions: For small and large $L_\bx$, the system should behave three dimensionally, and the region in between yields the non-trivial fractal dimension. If only the particle centers are considered, the algorithm simply counts the number of particles in the cluster for $L_\bx \lesssim d$ resulting in a horizontal line (dotted line).}
 \label{fig:ScalingConsideration}
\end{figure}

Only in the regime between these two limiting cases is it possible to
observe the fractal dimension with the box-counting method. Comparing
(\ref{eq:fracDimSmallL}) and (\ref{eq:fracDimLargeL}) shows that the
interesting range is proportional to $|\log \phi|$, which only depends
on the volume fraction, but not on the particular choice of the system
size. A schematic plot is given in Fig.~\ref{fig:ScalingConsideration}
where the number $N_\bx$ of boxes containing particles is plotted
against the edge length $L_\bx$ of a box.

For numerical reasons, it is very tedious to observe the expected
slope of $-3$ for {small} $L_\bx$, because of the vast amount of boxes
to account for. Since this regime is not relevant anyway, it has only
been investigated exemplarily and is reached for $L_\bx \lesssim 0.03
d\hc$. For all other runs we simplify the  algorithm and only use the
\emph{centers} of the particles, \emph{i.e.}~a box is only marked, if a
particle center is inside. With this definition the number of boxes
needed to cover the system for small box sizes $L_\bx < d\hc$ is just
the particle number $N$, resulting in a horizontal line on the left
side of the graph, instead of the slope $-3$ (dotted line in Fig.~\ref{fig:ScalingConsideration}).

\begin{figure}[h]
 \includegraphics[trim=0 50 0 50,width=.49\textwidth]{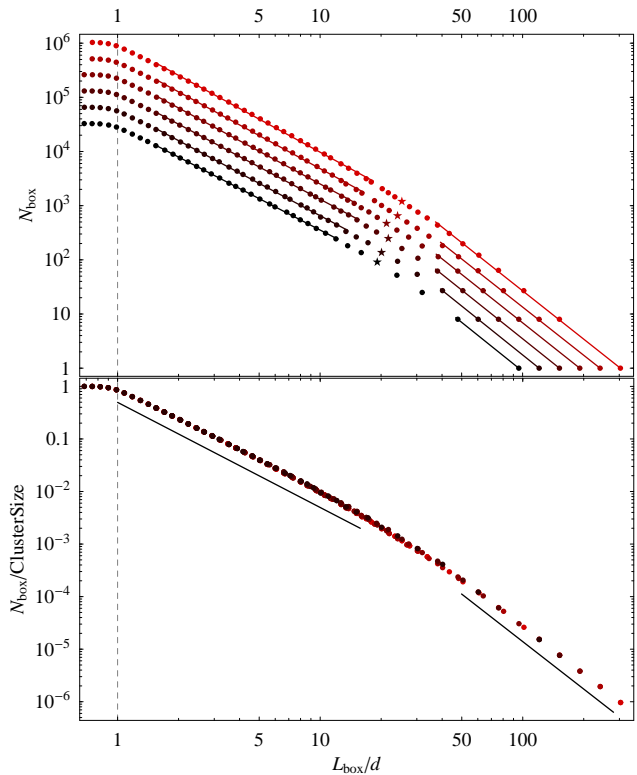}
  \centering
  \caption{(color online) Top: $N_\bx$ versus $L_\bx$ at time $t
    \approx 27 t_0$ and volume fraction $\phi = 1.95\%$ for the box
    counting algorithm; particle number is varied from bottom to top,
    according to $N = 32768, 65536, 131072,
    262144, 524288, 1048576$. The straight lines are fits to the
    data to the left and right of the cross-over point
    $L_\text{co} \approx 25d\hc$, which is also a fitting
    parameter and shown as a star ($\star$).  Bottom: As top, but
    $N_\bx$ normalized by cluster mass; the solid lines have 
    slopes $-2$ (left) and $-3$ (right); the
    vertical dashed line represents the particle size.} 
 \label{fig:HausdorffScale}
\end{figure}

Fig.~\ref{fig:HausdorffScale} (top) shows the outcome of the
box-counting algorithm, at a time $t\approx 27 t_0$, where roughly all
particles are inside the largest cluster. It yields the relation
between the box size $L_\bx$ and the number of boxes of that size,
needed to cover the cluster. The slope of that curve is the negative
fractal dimension. The result for different system sizes, but with the
same volume fraction are presented. As proposed, for all system sizes,
there is a cross-over point $L_\text{co}$, at which the slope
changes. On length scales between $d\hc$ and $L_\text{co}$, the
fractal dimension is roughly $2$ (the fits yield values between $1.92$
and $2.03$). Above $L_\text{co}$ the fractal dimension has a trivial
value of about $3$ (fit values between $2.95$ and $3.00$), which means
that on these large length scales all the boxes are filled and is
therefore an indication that the cluster is
system-spanning. 

In the lower half of Fig.~\ref{fig:HausdorffScale} we show the number
of boxes normalized by the cluster mass. The data collapse well onto a
single curve, obviously with the
same slopes. Here one can see very well that for systems with the same
volume fraction, the slopes as well as the cross-over point do not
depend on the absolute system size.

\begin{figure}[h]
 \includegraphics[width=.49\textwidth]{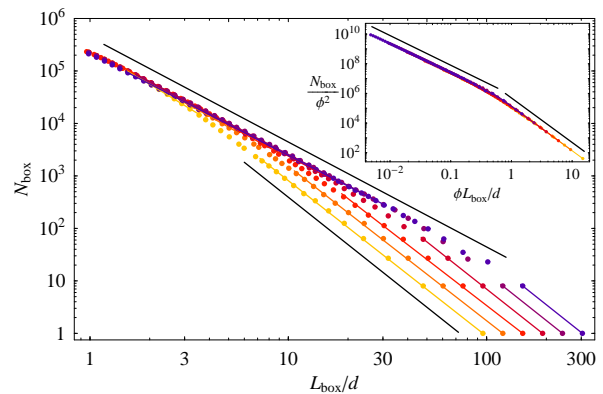}
  \centering
  \caption{(color online) Result of the box counting algorithm, as in
    Fig.~\ref{fig:HausdorffScale} at $t \approx 45 t_0$; the particle
    number is fixed $N = 262144$ and the volume fraction varies from
    left to right according to $\phi = 15.6\%, 7.81\%, 3.90\%, 1.95\%, 0.98\%,
    0.49\%$; the lines along the data points are the respective
    fits; the outer lines have slope $-2$ (top) and
    $-3$ (bottom).}
 \label{fig:HausdorffScaleDensity}
\end{figure}

Results of the box counting algorithm for different densities are
presented in Fig.~\ref{fig:HausdorffScaleDensity}. We only include
densities for which a spanning cluster has developed. The
\emph{slopes} of $-2$ and $-3$ in the two scaling regions are not
affected by the volume fraction $\phi$, but the size of the
non-trivial region (with $D_\text{f} \approx 2$) is seen to increase
significantly as the density decreases.  Even for the lowest density,
the size of the scaling region is less than 2 decades, which makes it
difficult to extract precise values for the fractal dimension. For the
three most dense systems, the scaling region is less than one
decade. As discussed earlier, this is an intrinsic feature of the
``high'' density systems, which can not be resolved by taking larger
systems ($N,L \to \infty$ with constant $\phi$). As the inset shows,
we can collapse all data on a single curve by rescaling $N_\bx$ with
$\phi^{-2}$ and $L_\bx$ with $\phi$ in agreement with the dependence of
the crossover length on $|\log \phi|$.

\subsubsection{Coordination number\label{sec:LCCoordinationNumber}}

\begin{figure}[h]
 \includegraphics[width=.49\textwidth]{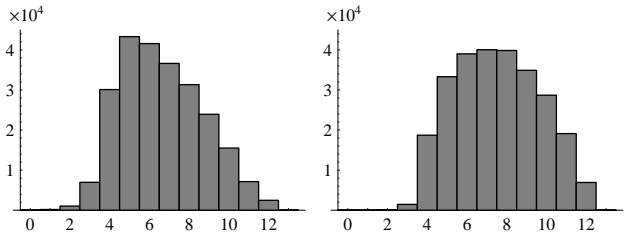}
  \centering
  \caption{Histogram of the coordination number for two different bond
    breaking distances $d_\crit = 1.01 d\hc$ (left) and $d_\crit = 1.07 d\hc$ (right); for both plots $N=262144$ and $\phi = 1.95\%$.}
 \label{fig:distr_coordNumber}
\end{figure}

Given the definition of a neighborhood relation (two particles are
neighbors if they have built a bridge and their kinetic energy is
not sufficient to break it), we can extract the average number of
neighbors of a particle, \emph{i.e.}~the average coordination number. 
In Fig.~\ref{fig:distr_coordNumber} we show  histograms for the
coordination number in the percolating cluster for two different bond
breaking distances. As one would expect, these distributions are
rather broad with coordination numbers between one and thirteen. The
smaller bond-breaking distance (left) gives rise to a more asymmetric
distribution with more weight for smaller coordination numbers.
  
\begin{figure}[h]
 \includegraphics[width=.49\textwidth]{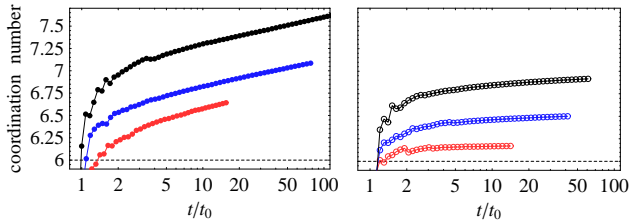}
  \centering
  \caption{(color online) Time evolution of the average coordination
    number of particles in the largest cluster ($N=262144$ and $\phi
    = 1.95\%$); the critical break-off distances are $d_\crit = 1.07
    d\hc, 1.035 d\hc$ and $1.01 d\hc$ from top to bottom;  thick film
    model (left) in comparison to thin film model (right).}
 \label{fig:dc_coordNumber}
\end{figure}

In Fig.~\ref{fig:dc_coordNumber}, we show the time evolution of the
average coordination number for different bond-breaking distances
$d_\crit$. After a strong increase at the time $t_0$,
the coordination number continues to grow slowly. This slow increase
is strongly suppressed in the thin film model (right) as compared to
the thick film model (left). Within the thin film model the slow
growth with time is further suppressed for decreasing bond-breaking
distance $d_\crit$.

\begin{figure}[h]
  \includegraphics[width=.49\textwidth]{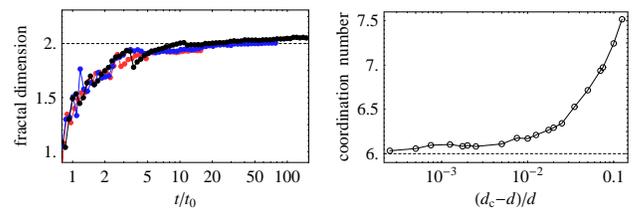}
  \centering
  \caption{Influence of the bond-breaking distance $d_\crit$ on the
    final value of the average coordination number for the thin film
    model (right, $\phi = 0.24\%$ and $N=10648$) and on the fractal dimension of the thick film model
    (left, system and colors as in Fig.~\ref{fig:dc_coordNumber}).}
 \label{fig:dc_coordThinFilm}
\end{figure}

As one can see in Figs.~\ref{fig:distr_coordNumber}, \ref{fig:dc_coordNumber}, the coordination
number becomes smaller for smaller $d_\crit$. This is reasonable,
because the particles can more easily collect neighbors for higher
$d_\crit$.  As $d_\crit \to d$ the average coordination number of the
thin film model approaches 6, which is the isostatic value. This is
demonstated in the right half of Fig.~\ref{fig:dc_coordThinFilm},
where we plot the asymptotic coordination number as a function of
$d_\crit$. Here the asymptotic value is taken, when $T< 0.06 \Delta E$
for the first time.

Naively one might expect that the increase of the coordination number
with larger $d_\crit$ is caused by a compactification and therefore
accompanied by an increase of the fractal dimension. However, as can
be seen in the left part of Fig.~\ref{fig:dc_coordThinFilm}, there is
no significant influenece of $d_\crit$ on the development of the
fractal dimension. Thus, we conclude that this compactification is
mostly occuring on the single particle length scale and therefore
increasing the average coordination number, but not influencing the
stucture on larger length scales \footnote{Note that there is also a
  very slight increase of the fractal dimension and therefore also a
  very slow compactification on larger length scales. However these
  effects are much less pronounced than the change of the coordination
  number.}.


\section{Conclusions\label{sec:conclusions}}

We have analysed a simple model of a wet granulate allowing for large
scale event driven simulations. A central feature of wet granulates is
the existence of an energy scale $\Delta E$ associated with the
rupture of a capillary bridge between two grains. This energy scale
has important consequences not only for the phase diagram
\cite{Fingerle2008b} but also for the free cooling dynamics investigated
in this paper.  The most important feature is a rather well defined
transition at a time $t_0$, when the kinetic energy $T$ of the
particles becomes equal to $\Delta E$.

For $t<t_0$ the particles are energetic enough to supply the bond
breaking energy $\Delta E$, so that very few collisions result in
bound pairs and most particles are unbound. Cooling is very effective
in this regime, but drastically different from a dry granulate.
Whereas in dry granulates the dissipated energy is proportional to the
energy of the colliding particles, in wet granulates the dissipated
energy is $\Delta E$, independent of the energy of the colliding
particles so that $\dot{T}\sim\sqrt{T}$. Consequently Haff's law does
not hold and is replaced by $T(t)=T(0)(1-t/t_0)^2$ for $t<t_0$. The
simulations are in very good agreement with this cooling law for $t<t_0$. 

For $t>t_0$, the kinetic energy of the particles is too small to
provide the bond breaking energy, so that larger and larger clusters
form. We call this regime the aggregation regime and analyse the
properties of the aggregates. For not too long times and sufficiently
small volume fractions, we observe flocculation characterized by
nonoverlapping, weakly interacting clusters. The fractal dimension of
the aggregates is approximately $D_f=2$.  The cluster size
distribution follows a simple scaling form, $N_m(t) \sim m^{-2}
f(m/\bar{m}(t))$, which has been applied successfully to different
aggregation models before. The increase of the typical cluster size
$\bar{m}(t)$ can be undestood by a simple scaling analysis: Assuming
that clusters irreversibly stick together when they hit upon each
other and that their radius $r$ grows with the number of particles $m$
like $r^{D_\text{f}} \sim m$, yields a cluster growth $\bar{m} \sim t^{{2
    D_\text{f}}/({3 D_\text{f} - 2 D + 2)}}$. This scaling relation
shows good agreement with the simulation for fractal dimension
$D_\text{f} = 2$.

At larger times, a spanning cluster forms, and a gelation transition is
observed for all finite volume fractions. At the gelation transition a
spanning cluster coexists with many small ones, wheras at very long
times almost all particles are connected to one large cluster.  On
the largest length scales the final cluster is no longer a fractal but
compact, as one would expect for a spanning cluster in the percolating
phase. On smaller length scales, however, we find fractal structures
with $D_f\approx 2$. The range where a nontrivial fractal dimension
can be observed increases with decreasing density as $|\log
\phi|$.

Even on the longest time scales, the temperature continues to
decay. In this regime the limiting process is the breaking of a bond.
The probability for this process becomes exponentially small $P_{bb}
\sim \sqrt{\Delta E/T}\, e^{-\Delta E/T}$ as the temperature goes to
zero. Hence the cooling law for high temperatures is replaced by
$\dot{T}\sim\ e^{-\Delta E/T}$ in very good agreement with the data.

Several extensions of our work might be interesting. So far we have
completely neglected all inelasticities except for the bond
rupture. One expects the collisions at the hard core to be dissipative
as they are in dry granular media. In the simplest model these could
be described by normal restitution. Furthermore real wet grains
experience frictional forces, coupling translational and rotational
motion of the grains\cite{Brilliantov07}. We are not aware of any such studies for wet
granulates.

\begin{acknowledgments}
We gratefully acknowledge financial support by the Deutsche Forschungsgemeinschaft (DFG) through Grant SFB 602/B6.
\end{acknowledgments}

\bibliographystyle{apsrev}
\bibliography{WetGranulates}

\end{document}